# Re-estimation of thermal contact resistance considering near-field thermal radiation effect

Yaoqi Xian[1], Ping Zhang[*,1], Siping Zhai[1], Peipei Yang[1], Zhiheng Zheng[2]

1 School of Mechanical and Electrical Engineering, Guilin University of Electronic Technology, No. 1 Jinji Road, Guilin, Guangxi 541004, China
2 College of Engineering, Shanghai Second Polytechnic University, Shanghai 201209, China
*corresponding author, E-mail: zp3631@gmail.com

**Abstract**

Thermal contact has always been a hot issue in many engineering fields and thermal contact resistance (TCR) is one of the important indicators weighing the heat transfer efficiency among the interfaces. In this paper, the contact heat transfer of conforming rough surfaces is theoretically re-estimated considering both the heat transfer from contact and non-contact regions. The fluctuational electrodynamics (an ab initio calculation) is adopted to calculate the thermal radiation. The contribution of contact regions is estimated by the CMY TCR model and further studied by modelling specific surfaces with corresponding surface roughness power spectrum (PSD). Several tests are presented where aluminum and amorphous alumina are mainly used in the simulations. Studies showed that there exists a significant synergy between the thermal conduction and near-field thermal radiation at the interface in a certain range of effective roughness. When the effective roughness is near to the scales of submicron, the near-field radiation effect should not be neglected even at room temperature.

**Nomenclature**

$A_a$ apparent area, m$^2$
$A_r$ real contact area, m$^2$
$C$ surface roughness power spectrum, m$^4$
$d$ separation, m
$D_f$ fractal dimension
$E$ Young's elastic modulus, N/m$^2$
$h$ height, m
$\hbar$ Planck constant divided by $2\pi$, J·s
$H$ microhardness, GPa
$i$ complex constant, $(-1)^{1/2}$
Im imaginary part
$k_B$ Boltzmann constant, J/K
$k_s$ harmonic mean thermal conductivity, W/mK
$k_0$ vacuum wavevector, rad/m
$k_\parallel$ wavevector parallel to the interface, rad/m
$k_\perp$ wavevector normal to the interface, rad/m
$k_q$ wavevector, 1/m
$m$ effective absolute surface slope
$P$ pressure, N/m$^2$
$q$ heat flux, W/m$^2$
$Q$ heat flow, W
$r_{ij}$ Fresnel reflection coefficients at interface i-j
R$_c$ thermal contact resistance, m$^2$K/W
R$_{cc}$ thermal constriction resistance, m$^2$K/W
$R_m$ thermal resistance of gap media, m$^2$K/W
$R_r$ thermal radiation resistance, m$^2$K/W
Re real part
$T$ temperature, K
$u_0$ length parameter, m
$v$ Poisson ratio
Y separation of mean planes, m

**Greek symbols**

$\gamma$ damping factor, s$^{-1}$
$\varepsilon_r$ dielectric constant
$\varepsilon_\infty$ high frequency dielectric constant
$\sigma$ effective roughness, m
$\Theta$ mean energy of a Planck oscillator, J
$\lambda$ relative separation
$\lambda_T$ characteristic wavelength, m
$\omega$ angle frequency, rad/s
$\omega_p$ plasma frequency, rad/s

**Subscripts/Superscripts**

*p* transverse magnetic, TM
*s* transverse electric, TE
*ω* monochromatic
*evan* evanescent wave
*LO* longitudinal optical
*prop* propagating wave
*TO* transverse optical

## 1. Introduction

Thermal contact resistance (TCR) is a core parameter weighing the heat transfer efficiency among the interfaces of different components. The phenomena of contact heat transfer are universal in many engineering fields such as aerospace, electronic packaging, cryogenics, and mechanical manufacturing [1-3]. Researches have pointed out that the thermal budget in thermal interface can account for half of the total in some microelectronic packages, which directly restricts the reliability, performance and lifetime of products [4]. So, it is crucial to qualitatively and quantitatively predicting the TCR in order to rationally carry out thermal management and design. As shown in Fig.1, the mechanism of the TCR is very complex and it is a result of triple interaction among geometry, mechanics and thermal [5]. In thermal transfer problem, there exist three modes including conduction of discrete point contact, convection of gap medium and radiation from a perspective of macroscopic scale, and they can also be interpreted as the transport in the form of energy carriers (i.e., phonons, electrons and photons) from a perspective of micro-nano scale [6, 7].

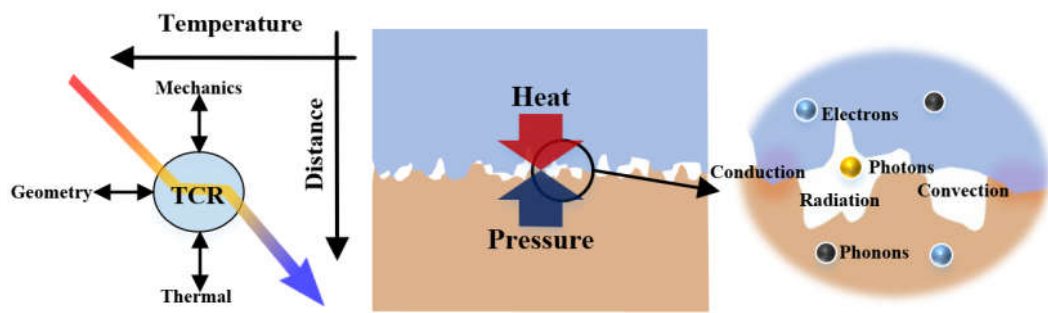

Fig. 1. Schematic for mechanism of thermal contact resistance

A large number of experiments have been implemented via diverse characterization methods to study the interfacial heat transfer that the interfacial separation is from atomic level to engineering application level (~μm) [2]. The researchers have investigated many factors with regard to the TCR, which includes the materials, the surface roughness and waviness, the interface temperature, the direction of the heat flux, the contact pressure, load cycle and the contact regions under different resolution [8-11]. However, the previous theoretical studies or models for the TCR usually ignore the influence of interfacial thermal radiation since the radiation effect governed by the Stefan-Boltzmann law on TCR is negligible compared to asperities contact heat conduction under the condition of low or room temperature. With the development of engineering technology, the minimum feature size of the device structure has been in the order of microns and continued to develop on the order of nanometers. At the micro-nano scale, the transport of matter and energy in any physical process takes place in confined microscopic geometries, such that the transport behavior of matter and energy exhibits different size effects at the macroscale [7]. In the late 1950s, Cravalho and Tien et al. [12] found a phenomenon that net radiative energy increases with decreasing separation and rapidly attenuated as the surface spacing increasing between two dielectrics, that is, near-field thermal radiation. The first correct calculation of near-field thermal radiation was done by Polder and Van Hove employing fluctuational electrodynamics [13]. More and more papers have reported that the near-field thermal radiative heat flux between two planar surfaces separated by a nanosized vacuum gap can exceed several orders of magnitude of the blackbody limit or even achieve a similar magnitude of the thermal conduction, on account of the electromagnetic evanescent waves, photon tunneling effects and excitation of surface polaritons [14-17]. As mentioned above, there are three main paths for heat to flow across the interface. The conduction and conduction with interstitial fluid have been proved to be significant for interfacial heat transfer in low or normal temperature. The thermal radiation in near field indicates that the energy transfer is mainly dependent on dielectric function of material and separation distance, and one should analyze it using fluctuational electrodynamics rather than conventional radiative transfer equation (RTE) based on particle features [17, 18]. In practice, the direct contact area between rough surfaces is much smaller than the apparent area, e.g., the diameter of the contact regions observed at atomic resolution may be of the order of ~1 nm [19, 20]. The enhancement of thermal radiation occurs when emitter and receiver are separated within characteristic wavelength ($\lambda_T$) obtained from Wien's displacement law. According to the formula ($\lambda_T \cdot T = 2898 \ \mu m \cdot K$), the characteristic wavelength of thermal radiation is about 10 μm at a temperature of 300 K. Obviously, the surface roughness of engineering interest is comparable or less than this magnitude. However, little attention has been paid to the near-field radiative effect on the TCR. The first work considering the near-field thermal radiation effect on the TCR was presented by Persson et al. [20, 21] using proximity approximation and their rough estimation of the MEMS device concludes that the non-contact contribution to heat transfer coefficient is larger than or of similar magnitude as the contribution from the area of real contact.

In this paper, we analyze the interfacial heat transfer of four metals including Al, Cu, Ag and Pb using the classical CMY TCR model as well as an ab initio calculation of thermal radiation to account for the heat transfer contribution of the non-contact regions. Among them, aluminum and amorphous alumina are chosen to carry out a coupling simulation for revealing the contribution between thermal conduction and near-field thermal radiation at a thermal contact interface. Finally, specific surfaces are constructed to further analyze and compare.

## 2. Methodology

### 2.1 Thermal contact resistance

To meet a wide range of thermal management applications, a number of experimental, analytical, numerical models to estimate the TCR have been developed. However, it is hard to derive a general precise predictive model due to the complexity of thermal joint or interface [22]. The TCR is defined as:

$$R_c = \frac{\Delta T}{q} = \frac{\Delta T}{Q/A_a} \tag{1}$$

where $\Delta T$ is the temperature drop at the interface, $q$ is the heat flux along the normal direction of the interface, $Q$ is the heat flow and $A_a$ is the apparent area of the interface. The reciprocal of the TCR is the thermal contact conductance (TCC).

Herein, we focus on the contact of metal surfaces of engineering interest that the roughness is mostly lower than 10 μm. Assuming the surfaces are microscopically rough and macroscopically conforming (i.e., nominally flat rough surfaces). Yovanovich et al. [23] developed a thermal contact correlation based on the classical Cooper-Mikic-Yovanocich (CMY) model considering both microscopic and macroscopic characteristics of a joint, which gives simple relationships for three measurable parameters of the contact surfaces: the geometric parameters, the mechanical parameters, the thermal parameters. The correlation has been verified comparing with plenty of experimental results and it is quite accurate for optically conforming surfaces [22, 24-26]:

$$R_{cc} = \frac{2\sigma\sqrt{2\pi}}{mk_s}\exp(\lambda^2/2)\left[1 - \sqrt{0.5 erfc(\lambda/\sqrt{2})}\right]^{1.5} \tag{2}$$

where $k_s = 2k_1k_2/(k_1 + k_2)$ is the harmonic mean thermal conductivity, $m = \sqrt{m_1^2 + m_2^2}$ is the effective absolute surface slope, $\sigma = \sqrt{\sigma_1^2 + \sigma_2^2}$ is the effective RMS surface roughness. $\lambda$ is a dimensionless parameter linking the geometry and the mechanics of the joint and it is defined as:

$$\lambda = \frac{Y}{\sigma} = \sqrt{2}erfc_{inv}(2P/H) \tag{3}$$

where $erfc_{inv}$ is the inverse transformation of the complementary error function, $P$ is the nominal pressure and $H$ is the microhardness of the softer metal.

Most surfaces produced by machining or grinding are Gaussian surfaces, that is, asperities are randomly distributed over the surface but isotropic and their profile heights obey the Gaussian distribution [22]. For Gaussian surfaces the empirical correlations to relate the effective RMS surface roughness, $\sigma$, to the effective absolute surface slope, $m$, is as follow [27, 28]:

$$m(\sigma) = \begin{cases} 0.124\sigma^{0.743}, & \sigma \leq 1.6 \text{ μm} \\ 0.076\sigma^{0.52}, & \sigma > 1.6 \text{ μm} \end{cases} \quad (4)$$

The distance between the mean planes can be derived via a detailed geometric analysis about interacting of two rough surfaces and the explicit correlation is expressed as [26]:

$$Y = 1.185\sigma \left[ -ln\left(3.132\frac{P}{H}\right) \right]^{0.547} \quad (5)$$

In this part, the classical thermal contact equations and other additional correlations are introduced to calculate the interface thermal resistance responsible for the sum of spot-to-spot contact in a vacuum condition. To be more explicit, Fig. 2 shows a schematic of a contact interface between conforming rough surfaces and the main geometry parameters within the equations above. Furthermore, the TCR can also be expressed using thermal resistance network:

$$\frac{1}{R_c} = \frac{1}{R_{cc}} + \frac{1}{R_m} + \frac{1}{R_r} \quad (6)$$

where $R_m$ and $R_r$ are resistance produced by gap media and radiation resistance respectively. Here, we only consider the situation of contact in vacuum that no interstitial fluids are present.

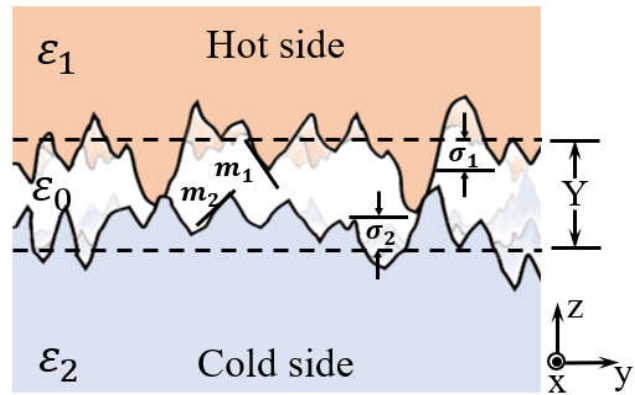

Fig. 2. Schematic of a contact interface between conforming rough surfaces

### 2.1.1 Surface roughness power spectrum

The CMY model above is used to preliminarily compare the heat transfer coefficient of contact conduction with radiative heat transfer of non-contact regions estimated by fluctuational electrodynamics. Then, we perform the further calculation via simulating specific surfaces with corresponding surface roughness power spectrum (PSD). The PSD can provide the characteristic information of a surface in wavevector space rather than real space, which is a more effective approach to describe the contact situation [20]. The PSD is defined as [29, 30]:

$$C(k_q) = \frac{1}{(2\pi)^2} \int d^2x \langle h(\pmb{x})h(\pmb{0})\rangle e^{-i\pmb{k}_q \cdot \pmb{x}} \tag{7}$$

where $\pmb{x} = (x,y)$ and $\pmb{k}_q = (k_{qx}, k_{qy})$. $h(\pmb{x})$ is the profile height measured from the mean plane. The sign $\langle ... \rangle$ represents the ensemble averaging. The mean square roughness amplitude as a function of the PSD can be determined by

$$\sigma^2 = \langle h^2 \rangle = 2\pi \int_{k_{q0}}^{k_{q1}} dk_q k_q C(k_q) \tag{8}$$

According to the contact mechanism proposed by Persson et al. [20, 31], in the case of low squeezing pressure, the heat transfer coefficient associated with the area of real contact is constructed:

$$h_c = \frac{Pk_s}{E^* u_0} \tag{9}$$

$$u_0 = \sqrt{\pi} \int_{k_{q0}}^{k_{q1}} dk_q k_q^2 C(k_q) \left( \sqrt{\pi \int_{k_{q0}}^{k_q} dk'_q k'^3_q C(k'_q)} \right)^{-1} \tag{10}$$

where $E^* = ((1-v_1^2)/E_1 + (1-v_2^2)/E_2))^{-1}$ is the effective Young's elastic modulus that is determined by the Young's elastic modulus and Poisson ratio of each solid. $u_0$ is a length parameter which is also determined from the PSD.

For the case of a self-affine fractal surface with a fractal dimension $D_f \leq 2.5$, Refs. [32, 33] give the relation between the interfacial separation ($d$) and the normal load ($P$):

$$d \approx \mu\sigma \ln\left(\frac{0.7493 k_{q0} \sigma E^*}{2P}\right) \tag{11}$$

where $\mu$ is a parameter that seems to relate to the surface roughness.

In addition, Majumdar et al. [34] have proved that the PSD of the equivalent surface is the sum of the power spectra of the individual surfaces:

$$C(k_q) = C_1(k_q) + C_2(k_q) \tag{12}$$

## 2.2 Radiative heat transfer

In previous research on the thermal interface, the radiation heat conductance is always neglected due to the complex manner of the bonding solids, and the contribution is relatively small compared with thermal conduction unless the surfaces have a high emissivity and are formed by rough, low-conductivity solids under light contact pressures [26]. However, previous estimations for the heat flux of interfacial radiation are based on the Stefan-Boltzmann law which is not included evanescent mode. Here, the radiative heat transfer is investigated employing an ab initio calculation based on the stochastic Maxwell equations and fluctuational electrodynamics [16,35]. The two components contacting with each other can be considered as semi-infinite (one-dimensional approximation), isotropic and non-magnetic. The spectral heat radiative flux calculation accounts for both the far- and the near-field effects [35, 36]:

$$q_{\omega,12}^{prop} = \frac{\Theta(\omega,T)}{4\pi^2} \int_0^{k_0} k_\parallel dk_\parallel \left[ \frac{(1-|r_{01}^s|^2)(1-|r_{02}^s|^2)}{|1-r_{01}^s r_{02}^s e^{2iRe(k_\perp)d}|^2} + \frac{(1-|r_{01}^p|^2)(1-|r_{02}^p|^2)}{|1-r_{01}^p r_{02}^p e^{2iRe(k_\perp)d}|^2} \right] \quad (13a)$$

$$q_{\omega,12}^{evan} = \frac{\Theta(\omega,T)}{\pi^2} \int_{k_0}^{\infty} k_\parallel dk_\parallel \, e^{-2iIm(k_\perp)d} \left[ \frac{Im(r_{01}^s)Im(r_{02}^s)}{|1-r_{01}^s r_{02}^s e^{-2Im(k_\perp)d}|^2} + \frac{Im(r_{01}^p)Im(r_{02}^p)}{|1-r_{01}^p r_{02}^p e^{-2Im(k_\perp)d}|^2} \right] \quad (13b)$$

where Eqs. (13a) and (13b) are the propagating and evanescent contributions to the monochromatic radiative heat flux between two bulks, respectively. $k_0^2 = k_\parallel^2 + k_\perp^2$ where $k_0$ is the wavevector in vacuum, $k_\parallel$ and $k_\perp$ are respectively the wavevectors parallel to and normal to the interface. $r_{ij}^{st}$ is the Fresnel reflection coefficients from medium $i$ to medium $j$ for $st$ polarization state ($s$ for TE and $p$ for TM). The mean energy of a Planck oscillator in thermal equilibrium at temperature $T$ of the source medium and angle frequency $\omega$, $\Theta(\omega,T)$, is given by

$$\Theta(\omega,T) = \frac{\hbar\omega}{\exp(\hbar\omega/k_B T) - 1} \quad (14)$$

where $\hbar$ is the Planck constant divided by $2\pi$ and $k_B$ is the Boltzmann constant.

The radiative resistance is defined as:

$$R_r = \frac{T_1 - T_2}{Q_r} = \frac{\Delta T}{\int_0^{+\infty} d\omega (q_{\omega,12}^{prop} + q_{\omega,12}^{evan})} \quad (15)$$

where $Q_r$ is the total radiative heat flux obtained by integrating over $\omega$ and it depends on the separation of two bulks, the temperature and the optical response of the materials.

### 2.2.1 Dielectric function and parameters of materials

The dielectric function $\varepsilon_r$ reflects the response of the materials to the external electric field. The dielectric function of metal and semiconductor can be described by the Drude model [14]:

$$\varepsilon_r(\omega) \equiv \varepsilon_r' + i\varepsilon_r'' = \varepsilon_\infty - \frac{\omega_p^2}{\omega^2 + i\gamma\omega} \tag{16}$$

where $\varepsilon_\infty$ denotes high-frequency contributions, $\omega_p$ is the plasma frequency, $\gamma$ is the damping frequency. Table 1 lists the parameters in regard to the dielectric function [37] and physical property [5, 24, 38] of the materials investigated commonly both in the TCR and near-field radiation.

In practice, the metal surface always has an oxide layer. The amorphous alumina is taken as an example to estimate the effect of a dielectric on near-field radiative heat transfer. The dielectric function of the film is described by the classical oscillator model:

$$\varepsilon_r(\omega) = \varepsilon_\infty + \sum_1^n \frac{S_n \omega_{T,n}^2}{\omega_{L,n}^2 - \omega^2 - i\omega\gamma_n} \tag{17}$$

where $n$ is the number of oscillators, $S_n$, $\omega_{T,n}$ and $\omega_{L,n}$ are the strength, TO mode frequency and LO mode frequency of the $n$th oscillator, respectively. The value of the parameters obtained from Ref. [39] is also list at table 1.

Table 1

The parameters of simulation materials

| Parameter<br>Material | $\varepsilon_\infty$<br>(1) | $\omega_p$<br>(rad/s)· $10^{16}$ | $\gamma$<br>(1/s)· $10^{14}$ | $H$<br>(GPa) | $k$<br>(W/m·K) |
|---|---|---|---|---|---|
| Al | 1 | 2.242 | 1.219 | 0.912 | 174 |
| Cu | 1 | 1.202 | 0.524 | 1.089 | 381 |
| Ag | 1 | 1.366 | 0.273 | 0.745 | 429 |
| Pb | 1 | 1.168 | 2.731 | 0.040 | 35.3 |
| $Al_2O_3$ | $\varepsilon_\infty$<br>(1) | $S_n$<br>(1) | $\omega_{T,n}$<br>(rad/s)· $10^{14}$ | $\gamma_n$<br>(1/s)· $10^{13}$ | $\omega_{L,n}$<br>(rad/s)· $10^{14}$ |
| n=1 | 2.8 | 3.75 | 0.795 | 3.196 | 1.012 |
| n=2 | 2.8 | 1.46 | 1.358 | 3.327 | 1.806 |

## 2.3 The coupling simulation strategy

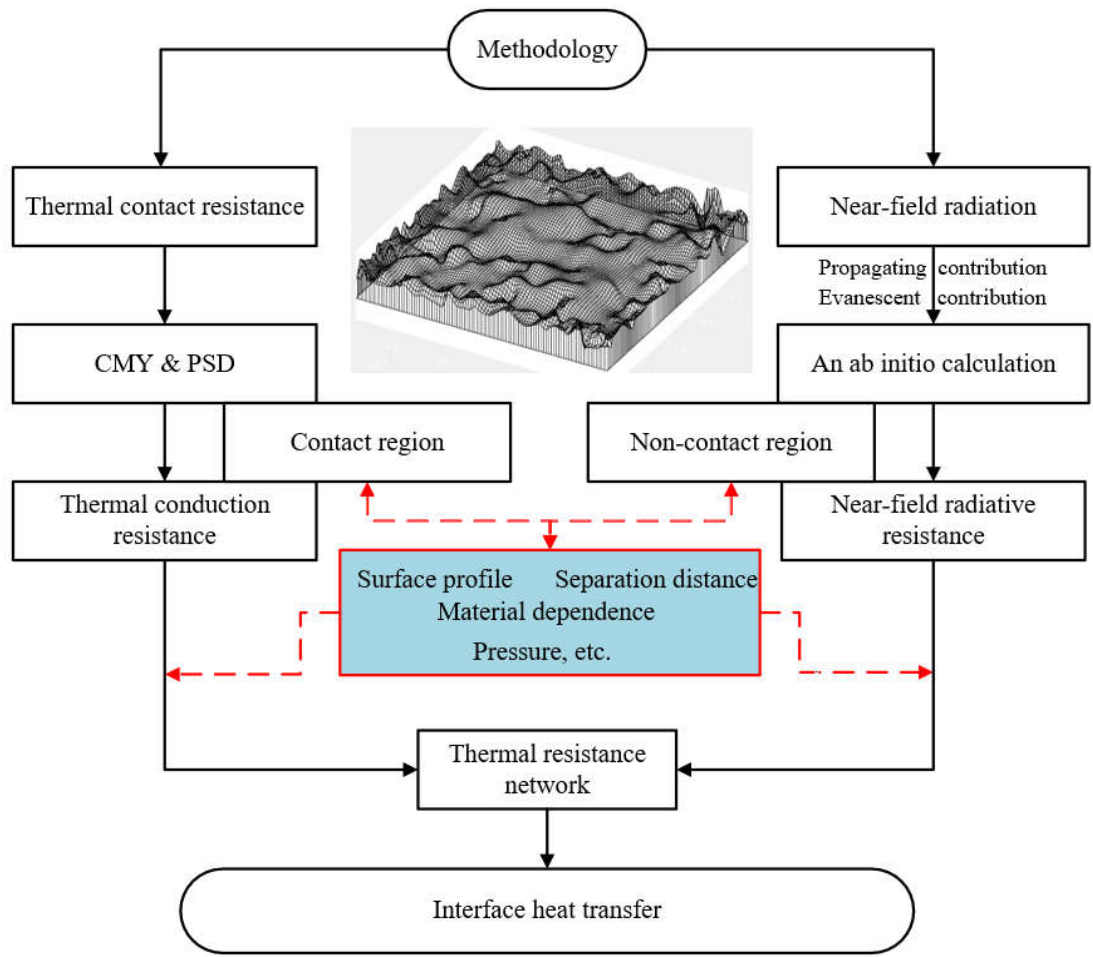

Fig. 3. Schematic of the coupling simulation strategy

Our intention is to investigate the influence of the near-field radiation at the interface when two bodies contacting together. CMY model and PSD method give the relation about conduction resistance accounting for the contact regions. The heat flux across these regions is determined by surface profile, roughness, apparent pressure, material, etc. On the other side, the radiative heat transfer accounting for non-contact regions are estimated using an ab initio calculation that it comprises the contribution of both the propagating and evanescent modes. In evanescent mode, the heat flux depends on material, temperature difference and separation distance. Obviously, some of the parameters of radiative heat flux are associated with the equations of thermal contact correlation, which means that there is a coupling relation. Fig. 3 describes the coupling simulation strategy to analyze interface heat transfer between conforming rough surfaces. The numerical algorithm of one-dimensional near-field thermal radiation problems is according to Francoeur et al. [36]. In this work, the integral terms are calculated using a composite Simpson's rule and the upper limit of integral for Eq. (13b) is set as $\pi * d_a^{-1}$ that $d_a$ is the distance between atoms. The convergence criteria of the integration are that the relative errors for propagating mode and evanescent mode are ≤1 and ≤0.01, respectively. The CMY model enables to roughly give us the insight that the near-field radiation plays an importance in which situation and then the specific surfaces are constructed to carry out a more rigorous investigation utilizing PSD.

## 3. Results and discussion

### 3.1 Heat transfer through the contact regions

The case of interface heat transfer in regard to the contact regions in vacuum is first investigated. In this area, heat flow transfers across the interface via microchannels constructed by the discrete asperities distributing on the upper and lower surfaces. If the heat leaves the half-space (body) through a small area (asperity), it forms an additional thermal resistance called constriction resistance due to the contraction of the flux lines. The converse is called spreading resistance. Therefore, there is a very complicated thermal resistance network in the interfacial heat transfer and the resistance network can also be determined by the macro measurable quantities in Sec. 2. Fig. 4 shows the curves of the TCR versus apparent pressure under different orders of magnitude of effective roughness of Al. Obviously, the TCR decreases with increasing pressure but increases with roughness. This tendency is in conformity with previous experimental investigation and theoretical models. In Fig. 5, we calculate the TCR of different metal materials with the roughness of a nanometer versus clamping pressure. It is of interest that in relatively high pressure lead demonstrates a well thermal dissipation at the interface compared with other three kind of metals although it possesses the lowest thermal conductivity. Because of low microhardness of the lead, the number of heat channels (i.e., microcontact spots) increases rapidly when the interface is subjected a larger load. The TCR of other three metals is mainly affected by their intrinsic thermal conductivities in the case of the same effective roughness and pressure. The Fig. 4 and Fig. 5 indicate a range of magnitudes of the TCR change with the effective roughness at different scales, which are compared with following thermal radiation resistance and analyze the role of near-field radiation at interfacial heat transfer.

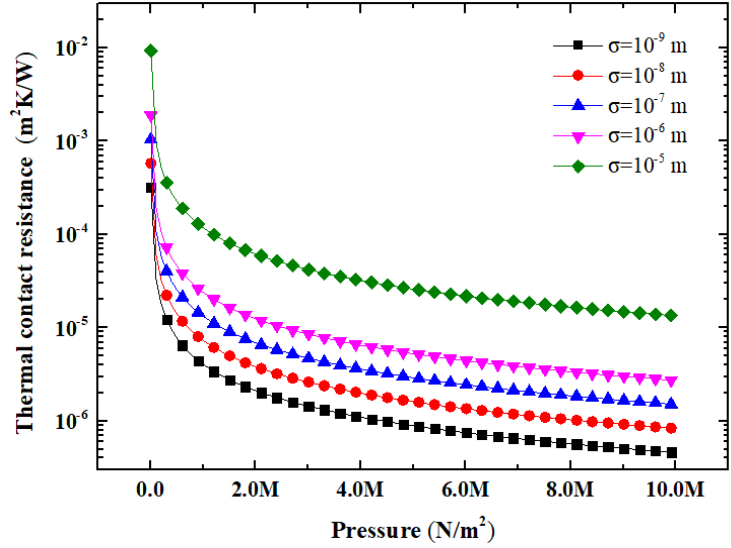

Fig. 4. Influence of clamping pressure on the TCR under different roughness

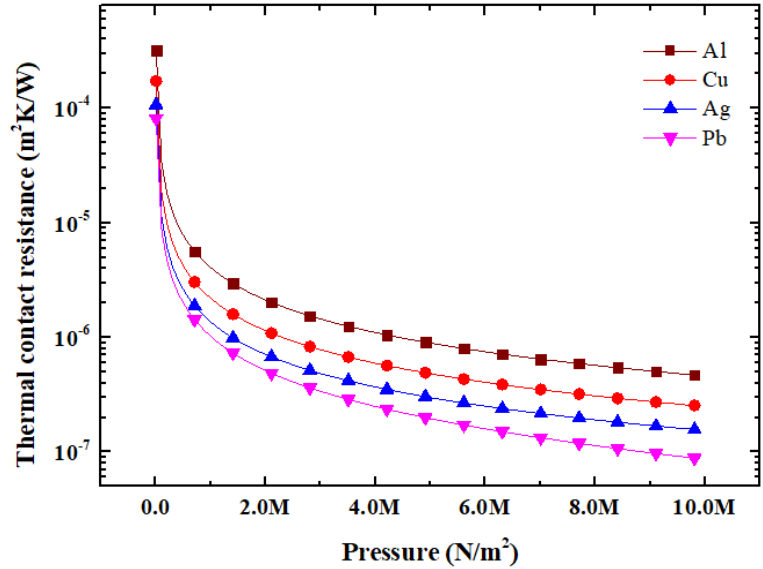

Fig. 5. Influence of clamping pressure on the TCR for different materials at 1 nm effective roughness

### 3.2 Heat transfer through the non-contact regions

According to the CMY TCR model, the ratio of real contact area to nominal area can be related to the apparent pressure and microhardness using the force balance condition:

$$\frac{A_r}{A_a} = \frac{P}{H} \tag{18}$$

In this work, the maximum pressure is set as 10.0 MPa and the minimum microhardness is 0.04 GPa so that the real contact area takes up ≤ 25% of the apparent area. For aluminum and copper, the ratio is less than 1%. Thus, the radiation effect is estimated by modeling the heat transfer across the micro-gap as equivalent to near-field thermal radiation between two infinite isothermal smooth plates, and the separation is assumed to equal to the distance between mean planes determined by effective roughness, pressure and microhardness. In the following simulation, the interfacial temperature difference used is 10 K, where the emitter (hot side) is 310 K and the receiver (cold side) is 300 K. The spectral radiative heat flux is enhanced in the near-field as shown in Fig. 6. At a separation of 1 nm, the spectral radiative heat flux comprised propagating mode and evanescent mode of the four metal materials are several orders of magnitude larger than the estimation for blackbodies (black line) that is independent with separation distance. The shapes of the curves are mainly determined by the dielectric function and the roughness effect on the near-field radiation is neglected.

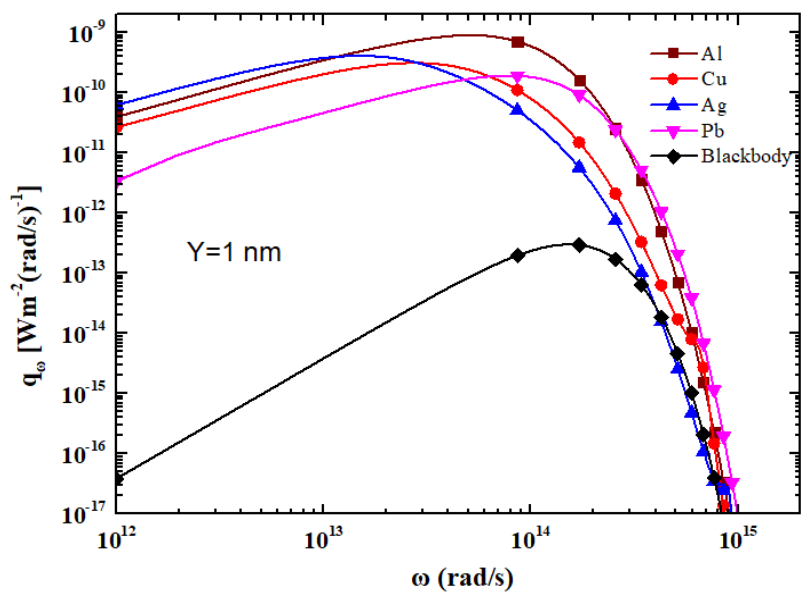

Fig. 6. Near-field spectral heat flux of different materials at a separation of 1 nm

The near-field thermal radiation between amorphous alumina (α-$Al_2O_3$) is also estimated for a rough approximation of the oxidation of aluminum. Note that the oxide layer is also simulated as bulk materials to simplify the structure and calculation. It can be seen in Fig. 7 that there are two peaks in the mid-infrared and at a separation of 1 nanometer the spectral radiative heat flux is seven orders of magnitude larger than blackbodies radiation and three orders of magnitude larger than the aluminum. The peaks account for the dielectrics that can support surface phonon-polaritons (SPhP). These surface electromagnetic waves are resonantly excited and provide the considerable contribution to the density of energy in the near-field [40]. An inset graph in the upper right corner is monochromatic evanescent component of the radiative heat flux per unit $k_{\parallel}$ at a separation of 100 nm. The two bright bands confined to a narrow range of frequencies in the inset correspond to the peaks nearby the angle frequency of $1.18\times10^{14}$ rad/s and $2.0\times10^{14}$ rad/s, respectively. When the separation achieves micrometer scale, this enhancement effect induced by the SPhP fades away. Fig. 8 illustrates the radiation resistance of Al and $Al_2O_3$. The overall radiative heat transfer performance of amorphous alumina is higher than that of aluminum.

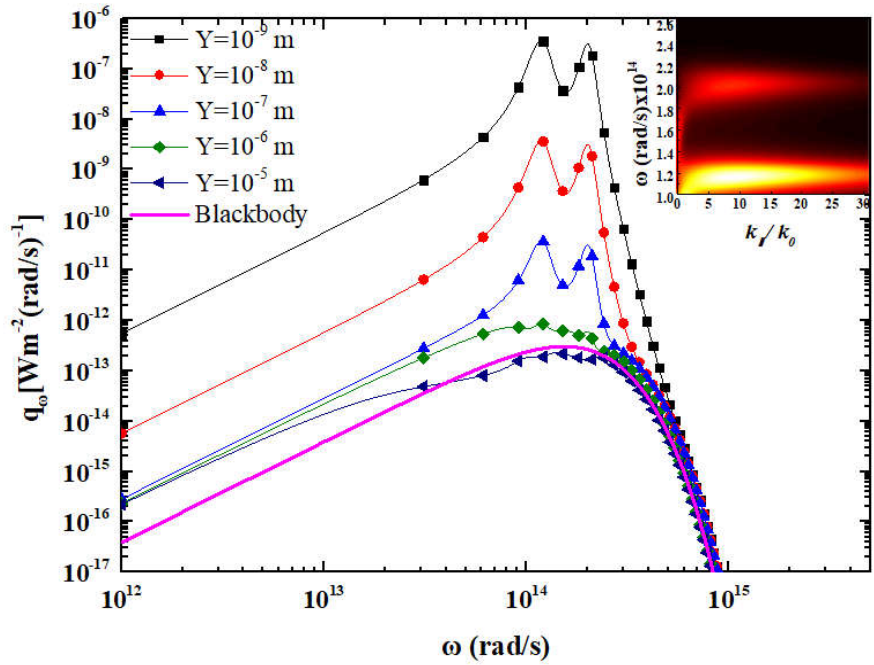

Fig. 7. Spectral radiative heat flux of the $Al_2O_3$ at different separation

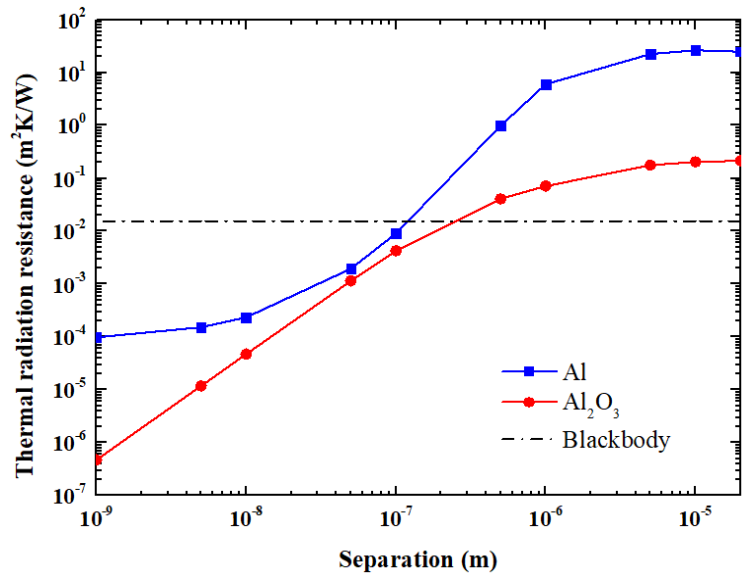

Fig. 8. Radiation resistance of the Al and $Al_2O_3$ at different scales

The interfacial temperature effect on the thermal radiation is also estimated at the separation of 0.1 μm and the cold side is fixed at 300 K. It can be seen in Fig. 9 the radiation heat transfer coefficient is approximately linearly related to the temperature difference in a specific range and the coefficient of the Al is not sensitive to the temperature changes compared with the Al$_2$O$_3$.

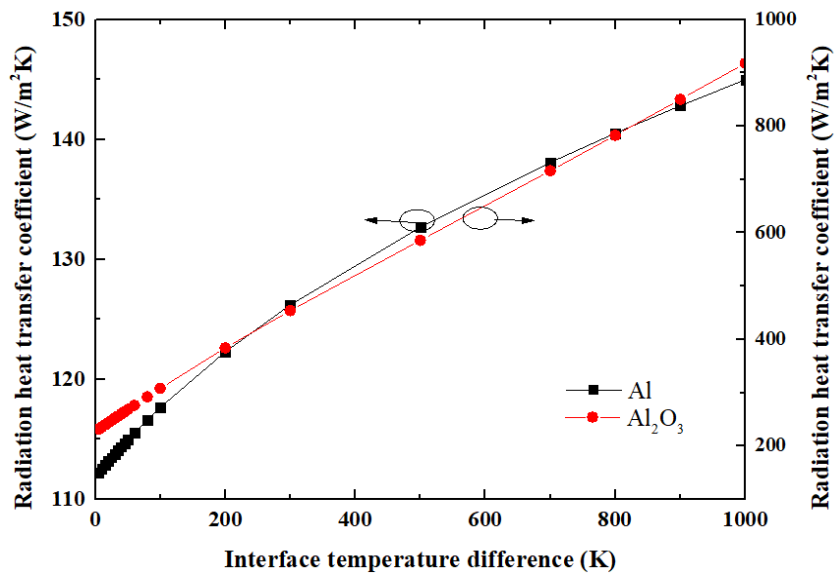

Fig. 9. Radiation heat transfer coefficient as a function of the interface temperature difference

### 3.3 Comparison and analysis of results

#### 3.3.1 Results based on CMY

As shown in Fig. 5 and Fig. 6, the Al possesses relative high TCR and near-field radiative heat transfer coefficient, which means that there is more likely to exist significant synergy at the interface. Furthermore, aluminum is a kind of frequently-used material in engineering. Therefore, the analysis and comparison are centered around the aluminum in this section. We first consider the heat transfer of both the thermal conduction and thermal radiation at the nanoscale as shown in Fig. 10. The effective roughness is fixed at 1 nm. As the applied pressure decreases, the mean plane spacing increases and the contact thermal conductance drops sharply. When the pressure is lower than 1 MPa, the near-field radiation of the Al$_2$O$_3$ is comparable to that of thermal conduction of the Al and gradually dominates the heat transfer. The inset with logarithmic plot of y axis shows the comparison between thermal conduction contribution and thermal radiation contribution of the Al, which enables to clearly display the case of extreme low pressure and large gap at the interface.

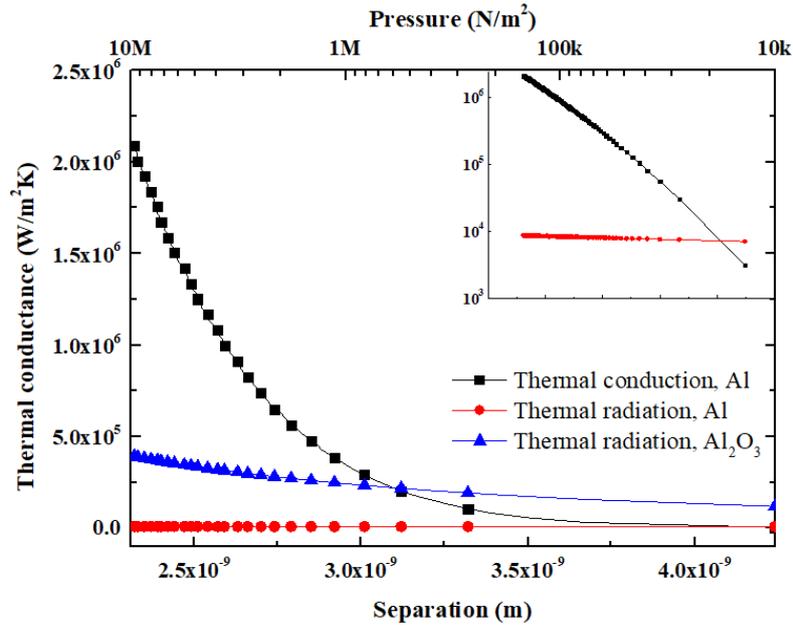

Fig. 10. Thermal conduction and thermal radiation at the nanoscale

When the pressure is fixed, the mean plane separation between two same solids and the thermal conduction resistance are mainly determined by the effective roughness. As shown in Fig. 11, the thermal conduction resistance and radiation resistance as a function of the scales of effective roughness at a low pressure of 1 kPa are investigated. Fig. 11a illustrates that the thermal radiation of the $Al_2O_3$ is dominant when the effective roughness is lower than 0.1 μm and produces the comparable effect to the thermal conduction at microscale. The thermal radiation of Al is also dominant in the nanoscale due to the contribution of the evanescent mode but it rapidly decreases and can be neglected at microscale.

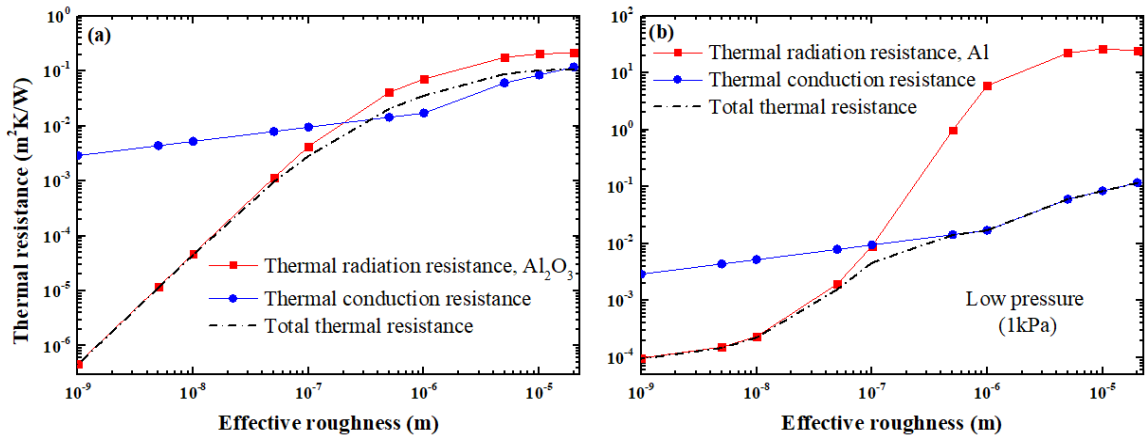

Fig. 11. The TCR versus the effective roughness at a pressure of 1 kPa

In order to clearly demonstrate the contribution of heat transfer from contact regions and non-contact regions, Fig. 12 shows the relative contribution from thermal radiation and thermal conduction as a function of the effective roughness based on the simulations of Fig. 11. From the tendency shown in Fig. 12a, the contribution of the thermal radiation of the $Al_2O_3$ may be larger than the thermal conduction again when the effective roughness (i.e., gap distance) becomes larger. Fig. 12b indicates that the near-field thermal radiation effect of pure metal aluminum should not be ignored at the effective roughness of submicron-scale, which approximately takes up 50% of the contribution to the interfacial heat transfer. At the separation of 100 nm, both the thermal radiation of the Al ($h_{Al}$=110.288 W/m$^2$K) and the $Al_2O_3$ ($h_{\alpha Al}$=235.569 W/m$^2$K) still exceed the blackbody radiation limit ($h_{rb}$=64.366 W/m$^2$K). It means that the near-field effect on the thermal radiation should be considered in the investigation of thermal contact.

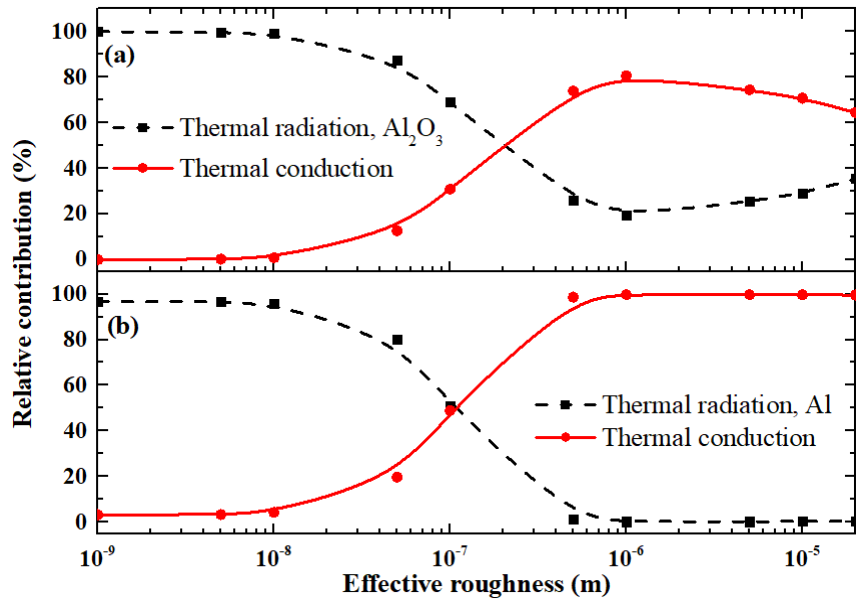

Fig. 12. Relative contribution from thermal radiation and thermal conduction as a function of the effective roughness

### 3.3.2 Results based on PSD

The analysis results from CMY model shows that when the effective roughness lower than submicron the impact of near-field thermal radiation is significant. Herein, we constructed three artificial randomly rough surfaces over an area of 10 μm × 10 μm and only consider the material of Aluminum. As shown in Fig. 13, a 3D fractal surface with effective RMS of 10 nm is modelled (Fig. 13a) and Fig. 13b shows 2D FFT of the surface topography. The symmetric power spectra indicate that the surface is isotropic [41]. The PSD profile of this surface has a cut-off wavevector ($k_{q0}$) of $10^7$ m$^{-1}$ and a slope of -3.203. It means that the fractal dimension $D_f$ is 2.4 since for a self-affine surface the PSD has the power-law behavior [29] $C(k_q) \sim k_q^{-2(4-D_f)}$. The wavevectors lower than the $10^7$ m$^{-1}$ determine the RMS. Fig 13d shows the height probability distribution of the surface, which is nearly Gaussian. The other two modeling surfaces (not shown) have the same properties but different effective RMS of 50 nm and 100 nm. We study such range of scale because the validity of fluctuational electrodynamics for sub-10-nm gap distances is still questionable.

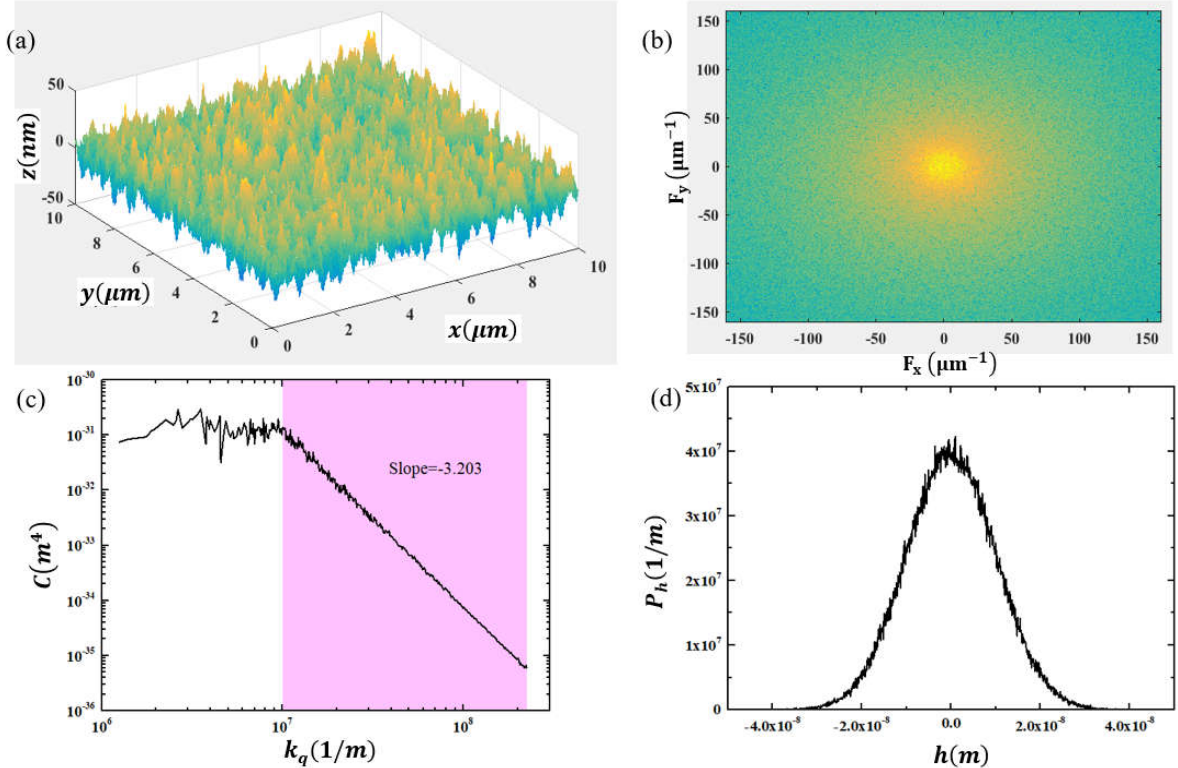

Fig. 13. Artificial randomly rough surface with effective roughness of 10 nm

It can be seen in Fig. 14 that the near-field radiation is seem to be weak at the effective roughness of 50 nm and 100 nm, which differs from the results based on CMY. The main reason is that the intensity of the near-field effect decays with the square of the distance and the separation distances used in near-field calculation are estimated via Eq. (11) and this equation assumes asperities undergone elastic deformation. Therefore, the real separation is several times larger than the effective roughness. The calculation results of gap distance of these three cases are 35.5 nm, 197.6 nm and 412.5 nm, respectively. The Eq. (9) and Eq. (10) are used to calculate the thermal conduction from contact regions.

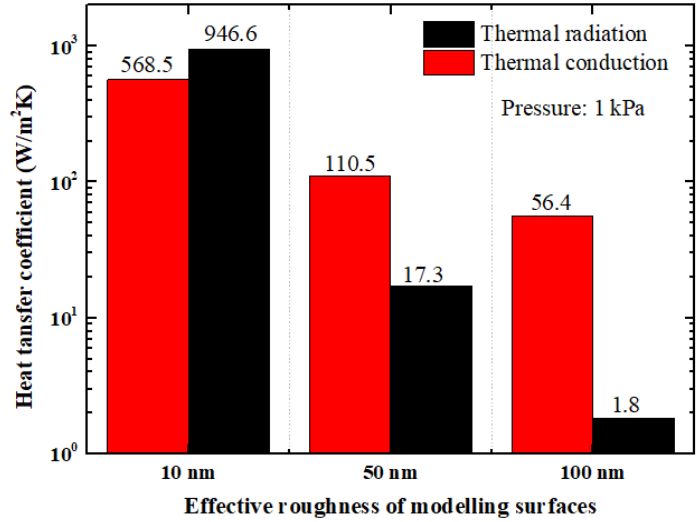

Fig. 14. Heat transfer coefficient comparison at three modeling surfaces

Fig. 15 shows the case of the modelling surface with an RMS of 10 nm. With the increase of squeezing pressure, the interfacial separation and thermal conduction resistance decrease observably but the near-field effect is not sensitive to the pressure in a small range. In this condition, the TCR is determined by the interaction of thermal conduction and near-field thermal radiation.

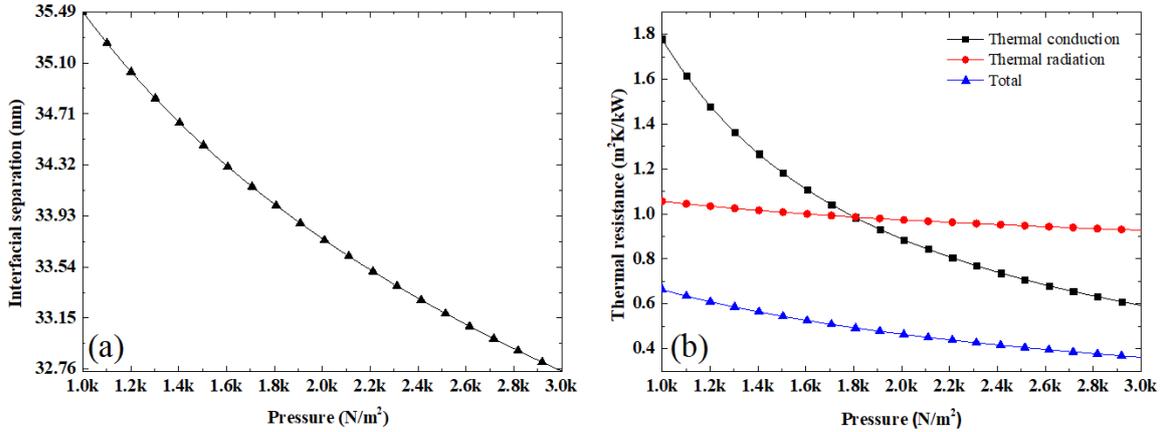

Fig. 15. The near-field thermal radiation effect on the TCR of modelling surface with an RMS of 10 nm

## 4. Conclusion

The role of near-field thermal radiation in thermal contact resistance has been preliminarily analyzed via the classical TCR CMY model and then further investigation are carried out utilizing PSD for specific rough surfaces. In most cases the separation between rough surfaces of engineering interest satisfies the condition of producing the near-field radiation effect at room temperature. The non-contact area accounts for the absolute dominance compared with the real contact area when two bodies come into bare contact. The simulation results show that the radiative heat flux of dielectric is obviously larger than that of metal because of the thermally excited SPhP in the near field. Therefore, the radiation effect of the dielectric on the TCR should be considered within a wider range of effective roughness than metal. At low pressure, a significant synergy exists between the thermal conduction and thermal radiation at the interface. When the effective roughness is lower than the scale of 0.1 μm, the near-field radiation in non-contact regions may make a difference to total heat transfer at room temperature. Simultaneously, the estimated radiative heat fluxes in such a spacing exceed the value predicted by the Stefan-Boltzmann law of blackbody. It easily achieves low pressure and low surface roughness conditions in microelectronics industry. The presented study is expected to provide a guidance for precisely predicting thermal contact problem.

**Acknowledgments**

The author(s) disclosed receipt of the following financial support for the research, authorship, and/or publication of this article: The authors acknowledge the financial support provided by National Natural Science Foundation of China (Project No. 51506033), Innovation Project of GUET Graduate Education (Grant No. 2016YJCX18), Guangxi Natural Science Foundation (Grant No. 2017JJA160108)).